\documentclass[10pt,twoside]{article}
\usepackage{own_sngl}
\usepackage[dvips]{graphics}
\usepackage{epsfig}
\usepackage{epsf}
\usepackage[figuresleft]{rotating}
\newcommand{\oversim}[2]{\protect{\mbox{\lower0.5ex\vbox{%
   \baselineskip=0pt\lineskip=0.2ex
   \ialign{$\mathsurround=0pt #1\hfil##\hfil$\crcr#2\crcr\sim\crcr}}}}} 
\newcommand{\simgreat}{\mbox{$\,\mathrel{\mathpalette\oversim>}\,$}} 
\newcommand{\simless} {\mbox{$\,\mathrel{\mathpalette\oversim<}\,$}} 
\slugcomment{\em MNRAS, accepted}
\lefthead{Kroupa}
\righthead{Clustered Star Formation and Galactic Disks}
\begin{document}
%
\title {Thickening of galactic disks through clustered star formation}
\author {Pavel Kroupa\\
\medskip
\small{Institut f\"ur Theoretische Physik und Astrophysik\\
Universit\"at Kiel, D-24098 Kiel, Germany}}
%
\begin{abstract}
\noindent 
The building blocks of galaxies are star clusters. These form with
low-star formation efficiencies and, consequently, loose a large part
of their stars that expand outwards once the residual gas is expelled
by the action of the massive stars. Massive star clusters may thus add
kinematically hot components to galactic field populations.  This
kinematical imprint on the stellar distribution function is estimated
here by calculating the velocity distribution function for ensembles
of star-clusters distributed as power-law or log-normal initial
cluster mass functions (ICMFs). The resulting stellar velocity
distribution function is non-Gaussian and may be interpreted as being
composed of multiple kinematical sub-populations.

The velocity-dispersion of solar-neighbourhood stars increases more
rapidly with stellar age than theoretical calculations of orbital
diffusion predict. Interpreting this difference to arise from star
formation characterised by larger cluster masses, rather than as yet
unknown stellar-dynamical heating mechanisms, suggests that the star
formation rate in the MW disk has been quietening down, or at least
shifting towards less-massive star-forming units.  Thin-disk stars
with ages 3--7~Gyr may have formed from an ICMF extending to very rich
Galactic clusters.  Stars appear to be forming preferentially in
modest embedded clusters during the past~3~Gyr.

Applying this approach to the ancient thick disk of the Milky Way, it
follows that its large velocity dispersion may have been produced
through a high star formation rate and thus an ICMF extending to
massive embedded clusters ($\approx10^{5-6}\,M_\odot$), even under the
extreme assumption that early star formation occurred in a thin
gas-rich disk. This enhanced star-formation episode in an early thin
Galactic disk could have been triggered by passing satellite galaxies,
but direct satellite infall into the disk may not be required for disk
heating.
\end{abstract}

\vskip 5mm
%

\section{INTRODUCTION} 
\label{sec:intro}
\noindent
Observations have shown that the Milky Way (MW) and other comparable
disk galaxies are composed of a number of more-or-less discrete
components. The broadest categories of these comprise the central
bulge with a mass $\approx10^{10}\,M_\odot$ and characteristic radius
of about 1~kpc, the Galactic spheroid (or stellar halo) with a mass
$\approx10^{8}\,M_\odot$ being mostly confined to within the solar
radius and also containing globular clusters that add up to a mass of
about $10^{6-7}\,M_\odot$, the embedded stellar and gaseous disk, and
a hypothesised extensive dark matter halo surrounding the lot and
extending to 20--200~kpc (Gilmore, Wyse \& Kuijken 1989; Binney \&
Merrifield 1998, hereinafter BM, for reviews). For the MW, the disk
can be subdivided into at least two components, namely the thin disk
with a mass of about $M_{\rm disk}=5\times10^{10}\,M_\odot$ with
exponential radial and vertical scale-lengths of
approximately~$h_R=3.5$~kpc and $h_z=250$~pc, respectively, and the
thick disk with roughly $h_{{\rm thd},R}=3.5$~kpc and $h_{{\rm
thd},z}\approx1000$~pc. Near the Sun, the thick disk comprises about
6~per cent of the thin disk mass (Robin et al. 1996; Buser, Rong \&
Karaali 1999; Vallenari, Bertelli \& Schmidtobreick 2000; Chiba \&
Beers 2000; Kerber, Javiel \& Santiago 2001; Reyl\'e \& Robin 2001),
so that the thick disk mass amounts to $M_{\rm thd} \approx (0.2-0.3)
\times M_{\rm disk}$. Mass-models of the MW, let alone of other disk
galaxies, remain rather uncertain though (Dehnen \& Binney 1998).

The structure of galaxies is linked to the physics of their formation
which is the topic of much on-going research (e.g. Chiba \& Beers
2000; Reyl\'e \& Robin 2001). Relevant to the long-term survival of
thin disks is understanding the origin of the Galactic thick disk.
This component is made up mostly of low-metallicity ([Fe/H]$\simless
-0.4$) stars that have a velocity dispersion perpendicular to the disk
plane of $\sigma_{z,{\rm obs}}\approx40$~pc/Myr, compared to the
significantly smaller $\sigma_z$ of the thin disk, which varies from
about 2--5~pc/Myr for the youngest stars to about~25~pc/Myr for stars
about 10~Gyr old (Fuchs et al. 2001). The classical work of Wielen
(1977) has shown that this increase can be understood as a result of
progressive heating of the thin disk population through a diffusive
mechanism.

The heating agent remains elusive though. This is nicely evident in
the work of Asiain et al. (1999), who find that spiral arms and a
central Galactic bar alone cannot account for the diffusion, and as
shown by Fuchs et al. (2001), scattering off molecular clouds also
cannot produce the necessary heating.  Jenkins (1992) also shows that
heating through spirals and molecular clouds and adiabatic heating
through mass growth of the Galactic disk do not lead to the observed
rise in velocity dispersion with stellar age.  Even worse, assuming
orbital diffusion does operate from some yet to be discovered
scattering agent(s), the large $\sigma_z$ for thick-disk stars cannot
be obtained from an extension of the Wielen-heating law. It follows
that the thick disk must have formed under different conditions. This
may be expected naturally, given that thick disk stars are not
substantially younger than the old halo stars, thus having been born
at the very beginning of the formation of the Galactic disk
(e.g. Gilmore \& Wyse 2001).

There are two broad classes of theories for the formation of a thick
disk component. One class of models involves the settling of an
initially hot proto-galactic gas cloud. A thick disk may form as a
result of dissipational settling to a thin disk (Burkert, Truran \&
Hensler 1992). According to the alternative hypothesis the MW formed
from many smaller components in a chaotic manner, leading first to the
formation of the spheroidal components, and a few~Gyr later, of the
thin gaseous disk. Perturbation of the early thin stellar disk through
continued accretion of dwarf galaxies may have lead to a thick disk
component, a scenario that has the advantage of not producing a
vertical metallicity gradient in the disk.  $N$-body computations,
however, may cast doubt on this scenario, because realistic dwarf
galaxies that have densities comparable to the parent galaxy are
likely to be destroyed before they venture close enough to the disk to
cause significant vertical heating (Huang \& Carlberg 1997; Sellwood,
Nelson \& Tremaine 1998; Velazquez \& White 1999), but the situation
is not clear since adequate computational resolution of gas-rich and
star-forming galactic disks is at present not possible.  The current
state of understanding concerning the MW thick disk is reviewed by
Norris (1999), who stresses that its origin remains unclear, and by
Gilmore \& Wyse (2001), who present the first results of an extensive
observational UK--Australian collaboration towards casting light on
this issue.

This paper addresses an additional mechanism that may be relevant to
the thickness of disks, and which is motivated by recent progress on
understanding star-cluster formation. Section~\ref{sec:csf} summarises
clustered star formation and details the calculation of the stellar
velocity distribution function. It also discusses known heating
mechanisms active in disk galaxies that increase the velocity
dispersion of stars with time. The large velocity dispersion of the
thick disk is considered in Section~\ref{sec:thickd} by studying the
velocity-field produced in a thin disk in which star-formation is
actively on-ongoing. The initial cluster mass function (ICMF) required
to give the observed velocity dispersion is constrained. Similarly,
the residuals of thin-disk age--velocity-\-dispersion data over
current understand of secular disk heating are analysed and linked to
the star-formation history in Section~\ref{sec:thind}.
Section~\ref{sec:disc} contains a discussion of the findings, and
concluding remarks follow in Section~\ref{sec:concs}.

\section{KINEMATICAL IMPLICATIONS OF CLUSTERED STAR FORMATION}
\label{sec:csf}
\noindent 
Stars are essentially never observed to form in isolation, but rather
in a spectrum of diversely-rich clusters. As outlined in Kroupa
(2001a) and Boily \& Kroupa (2001), a star-cluster forms stars over a
time period $\tau_{\rm clf}\approx1-2$~Myr in a contracting
gas-cloud. Because the collapse time of an individual proto-star takes
of the order of 0.1~Myr$\,\ll\tau_{\rm clf}$, each proto-star
decouples dynamically from the gas and has enough time to add to the
growing star$+$gas system that is, consequently, approximately in
virial equilibrium at any time provided the proto-cluster crossing
time (eqn~1 in Kroupa 2001a) is shorter than $\tau_{\rm clf}$. This is
true for cluster masses $M_{\rm cl}\simgreat 500\,M_\odot$ and cluster
radii of 1~pc. In sufficiently rich clusters, O~stars terminate
star-formation, and a large proportion of the cluster stars become
unbound as a result of rapid expulsion of the remaining gas and the
intrinsically low star-formation efficiency, $\epsilon=M_{\rm
cl}/(M_{\rm cl} + M_{\rm gas}) \simless 0.4$, where $M_{\rm cl}$ and
$M_{\rm gas}$ is the mass in stars and gas, respectively, before
gas-expulsion. Observations of very young clusters support this
scenario. For example, the Orion Nebula Cluster is at most~2~Myr and
probably only a few $10^5$~yr old, and already largely void of
gas. Similarly, the massive 30~Doradus cluster in the Large Magellanic
Cloud is not much older and has already removed its gas.

Direct computations of these processes using the Aarseth state-of-the
art direct $N$-body code are now available. Beginning with $N=10^4$
stars in an Orion-Nebula-\-Cluster-\-like configuration, and assuming
$\epsilon=0.33$ with the residual gas being blown out faster than the
cluster's dynamical time, Kroupa, Aarseth \& Hurley (2001) find that
about~2/3 of the cluster stars are lost and form a rapidly expanding
stellar association, while a Pleiades-like cluster condenses as a
nucleus near the origin of the flow. The relative number of massive
stars remaining in the core that forms the cluster, and those that are
lost, depends on the initial concentration of the cluster prior to gas
expulsion, and whether the massive stars form near the centre of the
cluster. This scenario is supported by the fact that the velocity
dispersion in the Orion Nebula Cluster is super-virial (e.g. Jones \&
Walker 1988), so that it is probably expanding now (Kroupa, Petr \&
McCaughrean 1999; Kroupa 2000; Kroupa, Aarseth \& Hurley 2001).
Especially striking in this context is the very recent measurement of
the velocity dispersion of stars in the 30~Doradus cluster by Bosch et
al. (2001), who find $\sigma\approx35$~pc/Myr. This is too large for the
cluster's mass, and the authors attribute the surplus kinetic energy
as being due to binary-star orbital motion and a binary proportion
among the massive stars of 100~per cent. The notion raised here is
that the velocity dispersion may also appear inflated due to recent
gas blow-out.  This notion that star clusters form as the nuclei of
expanding OB associations is also (retrospectively) supported by the
distribution of young stars around the 30~Doradus cluster suggesting
it is the core of a stellar association (Seleznev 1997), and the
likely association of the $\alpha$~Persei cluster with the Cas--Tau
OB~association noted by Brown (2001), and other moving groups with
star-clusters (Chereul, Cr\'ez\'e \& Bienaym\'e 1998; Asiain et
al. 1999). Furthermore, Williams \& Hodge (2001) note that most of the
young clusters in M31 are located within large OB associations.

The unbound population expands approximately with a one-dimensional
velocity dispersion, $\sigma$, typical of the pre-gas-expulsion
cluster$+$gas mixture. Neglecting factors of the order of one,
\begin{equation}
\sigma \approx \sigma_{0,{\rm cl}} = \sqrt{{G\,M_{\rm cl} \over \epsilon R_0}},
\label{eqn:sigma}
\end{equation}
where $G=0.0045\, {\rm pc}^3/(M_\odot\, {\rm Myr}^2)$ is the
gravitational constant and $R_0$ the characteristic embedded-cluster
radius.  It becomes immediately apparent that $\sigma_{0,{\rm
cl}}=40$~pc/Myr (1~km/s$\,\approx\,$1~pc/Myr) corresponds to
\begin{equation}
{M_{\rm cl} \over \epsilon R_0} = 10^{5.5}\,M_\odot/{\rm pc},
\label{eqn:typm}
\end{equation}
coming close to a typical globular cluster mass if
$\epsilon\,R_0=0.3$~pc, which is typical for young embedded clusters
that have characteristic radii $R_0\approx1$~pc ($R_0=1$~pc is assumed
in what follows, unless stated otherwise; examples of results with
$R_0=5$~pc are given in Fig.~\ref{fig:survad}). Most of the variation
of $\sigma$ stems from variations of the embedded cluster mass which
spans many orders of magnitude in contrast to the typical radii of
modest embedded clusters and very young massive clusters, all of which
are more concentrated than a few~pc. For example, local modest
embedded clusters have $R_0\simless 1$~pc (Kaas \& Bontemps 2001; Lada
\& Lada 1991), while the rich Orion Nebula Cluster has $R\approx2$~pc
(Hillenbrand \& Hartmann 1998).  Young massive clusters in external
galaxies have $R\approx2-5$~pc (e.g. R136 in the Large Magellanic
Cloud: Massey \& Hunter 1998; Sirianni et al. 2000) and in massively
interacting gas-rich galaxies, the Antennae, the young massive
clusters remain unresolved with $R\simless5$~pc (Whitmore 2001). These
clusters are, however, already void of their gas despite being less
than a few Myr old, so they should have already expanded
($R>R_0$). This is also true for the somewhat older clusters in a
sample of early-type galaxies compiled by Larsen et al. (2001)
($R\simless 4$~pc), and in M31 Williams \& Hodge (2001) note cluster
radii $R\simless5$~pc for cluster ages between about~10 and 200~Myr.

Expansion also implies that the observed velocity dispersion in
clusters is smaller than $\sigma_{0,{\rm cl}}$. For example, Ho \&
Filippenko (1996) measure a line-of-sight velocity dispersion of
about~11--16~pc/Myr for two extragalactic young massive star clusters.
These have inferred masses of about $10^4\,M_\odot$ and are 10--20~Myr
old. Smith \& Gallagher (2001) find~13~pc/Myr for a young massive
cluster ($60\pm20$~Myr old) in~M82, and deduce its mass to be about
$10^6\,M_\odot$. If the present notion is correct and if
$\epsilon=0.3$, then the deduced cluster masses correspond to only
about 30~per cent of their birth stellar masses ($M_{\rm cl}$).

Star-clusters with masses indicated in~(\ref{eqn:typm}) are observed
to form in profusion whenever gas-rich galaxies interact or are
perturbed (Lancon \& Boily 2000; Larsen 2001 for overviews). It is
thus feasible that eqn~\ref{eqn:typm} may indicate that the thick disk
resulted from massive clusters forming in a thin disk, rather than
being produced through orbital scattering by a merging satellite
galaxy.  Perturbation of an early thin and gas-rich MW disk by a
passing satellite may suffice to trigger the required star formation.

\subsection{The velocity distribution function}
\label{sec:veldistr}
\noindent 
To address the above question, the distribution function of stellar
velocities resulting from a burst of star formation needs to be
estimated. Attention is focused on velocities perpendicular to the
Galactic disk because motions in the plane are much more difficult to
handle analytically since they depend on the evolving mass
distribution throughout the entire MW. An extension to include the
velocity ellipsoid would go beyond the scope of this scouting work,
but will be addressed in future contributions.

As an ansatz the distribution of stellar velocities in one-dimension
(perpendicular to the Galactic disk, $v_z$) in the cluster prior to
gas-expulsion is assumed to be Gaussian (i.e. a one-dimensional
Schwartzschild distribution),
\begin{equation}
{\cal G}({\sigma_z},\overline{v_z}) = { 1\over \sqrt{2\,\pi} \,\, \sigma_z} 
              e^{ - {1\over2} ({v_z-\overline{v_z}\over \sigma_z})^2 },
\label{eqn:gausvd}
\end{equation}
$\overline{v_z}$ being the centre-of-mass velocity of the cluster, and
$\int_{-\infty}^{+\infty}{\cal G}\,dv_z = 1$. Gas expulsion leads to
isotropic expansion, and the velocity dispersions in galactic radial
and tangential directions can be linked to the z-component via the
epicyclic approximation if sufficiently small compared to the circular
velocity about the galaxy (Binney \& Tremaine 1987, hereinafter
BT). Large expansion velocities would require detailed orbit
integration in a self-consistent potential, which goes beyond the aim
of the present study.  After the gas is expelled from the cluster the
major part of the stellar population is assumed to expand freely
conserving this distribution but with a velocity dispersion somewhat
reduced due to self-gravity, leaving behind the nucleus that forms the
bound cluster (Kroupa et al. 2001). The cluster fills its tidal
radius and has a much smaller velocity dispersion than $\sigma_{0,{\rm
cl}}$, but analytical quantification is difficult since the processes
operating in shaping the cluster are complex (three-body and four-body
encounters re-distributing kinetic energy into potential energy,
stellar evolution, tidal field).

The stellar velocity distribution function after gas expulsion for an
ensemble of co-eval identical clusters with mass $M_{\rm cl}$ becomes
\begin{equation}
{\cal V}(v_z,M_{\rm cl}) 
         = \kappa\,{\cal E}(v_z,M_{\rm cl}) + 
           \left(1-\kappa\right)\,{\cal C}(v_z),
\label{eqn:truevd}
\end{equation}
where $\kappa(\epsilon)$ is the fraction of stars expelled after the
gas is thrown from each cluster. The distribution function of the
expanding stars is
\begin{equation}
{\cal E}(v_z,M_{\rm cl}) = {\cal G}(\sigma_{{\rm exp},z},\overline{v_z}=0),
\label{eqn:exp}
\end{equation}
since the centre-of-mass of the ensemble is stationary, and where 
\begin{equation}
\sigma_{{\rm exp},z}^2(M_{\rm cl}) = 
           {G\,M_{\rm cl} \over \epsilon R_0}\,\zeta + \sigma_{0z}^2,
\label{eqn:sigmaexp}
\end{equation}
is the velocity variance of the freely expanding populations, with
$\sigma_{0z}$ being the cluster--cluster velocity dispersion resulting
from a cluster centre-of-mass velocity distribution that is assumed to
be Gaussian (eqn~\ref{eqn:gausvd}). In the present thin MW-disk
molecular clouds move relative to each other with a velocity
dispersion of about~5~pc/Myr (e.g. Jog \& Ostriker 1988), so that
$\sigma_{0z}=5$~pc/Myr is adopted in what follows.  Gravitational
retardation of the expanding population is approximated through the
reduction factor, $\zeta=1-\epsilon$, which comes from the loss of
kinetic energy, as the fraction, $\kappa\,M_{\rm cl}$, of the cluster
expands out of the cluster potential well to infinity ($\sigma_{{\rm
exp},z}^2 - \sigma_{0z}^2 = \sigma_{\rm 0,cl}^2 - G\,M_{\rm cl}/R_0$,
where $\sigma_{\rm 0,cl}^2$ is the velocity variance in the cluster
prior to gas expulsion, eqn~\ref{eqn:sigma}).

In what follows it is assumed that the clusters live short lives
relative to the age of their galaxy due to evaporation through
two-body relaxation; any remaining clusters contributing an
insignificant amount of stars. The clusters ultimately leave an
extremely long-lived cluster remnant consisting of a strongly
hierarchical multiple star system (de la Fuente Marcos 1997; de la
Fuente Marcos 1998).  The velocity distribution function of stars
remaining in the cluster remnants and in the resulting tidal tails and
moving groups is summarised as
\begin{equation}
{\cal C}(v_z) = {\cal G}(\sigma_{{\rm cl},z},\overline{v_z}=0).
\label{eqn:cl}
\end{equation}
The velocity variance of the stars that ultimately leak out of the
bound clusters that form as the nuclei of the expanding associations
is approximated here by
\begin{equation}
\sigma_{{\rm cl},z}^2 = 2^2\,{\rm pc/Myr} + \sigma_{0z}^2,
\label{eqn:sigmacl}
\end{equation}
taking the velocity dispersion in the cluster remnant and of its
evaporated stars to be 2~pc/Myr.

Note that $\int_{-\infty}^{+\infty}{\cal V}\,dv_z = 1$, and
$\kappa=1-\epsilon$ with $\epsilon=0.33$ is adopted based on the
results of Kroupa et al. (2001) and discussion therein.  The
star-formation efficiency, $\epsilon$, may, at most, be a weak
function of $M_{\rm cl}$, possibly achieving~0.5 for massive clusters
(Tan \& McKee 2001; Matzner \& McKee 2000).

Each cluster contributes $M_{\rm cl}/m_{\rm av}$ stars to the field
population, where $m_{\rm av}=0.4\,M_\odot$ is assumed to be the
invariant average stellar mass (Kroupa 2001b and Reyl\'e \& Robin 2001
note possible evidence for a weak dependency on metallicity). The
total number of stars that result from a burst of star formation is
thus
\begin{equation}
N = {1\over m_{\rm av}} \int\limits_{M_{\rm cl,min}}^{M_{\rm cl,max}} 
    M_{\rm cl} \, \xi(M_{\rm cl})\,dM_{\rm cl} = 
    {\eta'\,M_{\rm disk}\over m_{\rm av}},
\label{eqn:nstars}
\end{equation}
where $\xi(M_{\rm cl})\,dM_{\rm cl}$ is the number of clusters with
masses in the range $M_{\rm cl}$ and $M_{\rm cl}\,dM_{\rm cl}$, and
$\eta'$ is the fraction of the total galactic disk mass in this
population of clusters.

The initial cluster mass function (ICMF), $\xi(M_{\rm cl})$, is
extremely hard to infer from the observed distribution of cluster
luminosities, because young clusters rapidly evolve dynamically as
well as photometrically, and since the light need not follow mass by
virtue of mass segregation.  Stellar-dynamical evolution of young or
forming clusters stands only at the beginning stages (Kroupa 2000;
Kroupa et al. 2001), while significant uncertainties in stellar
evolution continue to hamper the interpretation of the stellar
population in clusters, let alone of integrated cluster luminosities
and colours of unresolved clusters (Santos \& Frogel 1997).  Some
progress is evident though, and for example Whitmore et al. (1999),
Elmegreen et al. (2000), and Maoz et al. (2001) report, for different
star-forming systems, power-law ICMFs,
\begin{equation}
\xi_{\rm P}(M_{\rm cl}) = \xi_{\rm 0P}\, M_{\rm cl}^{-\alpha},
\label{eqn:pol}
\end{equation}
with $\alpha=1.5-2.6$. Old globular clusters, however, are typically 
distributed normally in log$_{10}M_{\rm cl}\equiv lM_{\rm cl}$, which is also
the distribution of young clusters reported by Fritze--von
Alvensleben (2000) for the interacting Antennae galaxies,
\begin{equation}
\xi_{\rm G}(lM_{\rm cl}) = { \xi_{\rm 0G} \over \sqrt{2\,\pi} \,\,
              \sigma_{\rm lMcl} } e^{ - {1\over2} ( {lM_{\rm
              cl}-\overline{lM_{\rm cl}} \over \sigma_{\rm lMcl}})^2},
\label{eqn:gaus}
\end{equation}
where $\overline{lM_{\rm cl}} \approx 5.5$ is the mean log-mass and
$\sigma_{\rm lMcl}\approx0.5$ is the standard deviation in log-mass
($M_{\rm cl}$ in units of $M_\odot$), and $\xi_{\rm G}(lM_{\rm
cl})\,dlM_{\rm cl}$ is the number of clusters with log-mass in the
interval $lM_{\rm cl}$ to $lM_{\rm cl}\,dlM_{\rm cl}$.

From eqn~\ref{eqn:nstars},
\begin{equation}
\xi_{\rm 0P} = m_{\rm av}\,N \left({2-\alpha \over 
               M_{\rm cl,max}^{2-\alpha} - M_{\rm cl,min}^{2-\alpha}}\right),
~~~~~~\alpha\neq2, 
\label{eqn:polnorm}
\end{equation}
and
\begin{equation}
\xi_{\rm 0G} = m_{\rm av}\,N \,\,\, 10^{-\overline{lM_{\rm cl}}}
               e^{-{1\over2}\,({\rm ln}10\,\sigma_{\rm lMcl})^2}
\label{eqn:gausnorm}
\end{equation}
for integration over $lM_{\rm cl}$ with $lM_{\rm cl,min}=-\infty$ to
$M_{\rm cl,max}=+\infty$, remembering that $10^x=e^{x\,{\rm ln}10}$.

The distribution function of stellar velocities, $D(v_z)$, that results from
the formation of an ensemble of star clusters, becomes
\begin{equation}
D_{\rm P}(v_z; M_{\rm cl,max},\alpha) = {1\over m_{\rm av}}\,
         \int\limits_{M_{\rm cl,min}}^{M_{\rm cl,max}} 
         M_{\rm cl}\,\,\xi_{\rm P}(M_{\rm cl})\,\,
         {\cal V}(v_z,M_{\rm cl}) 
         \,\, dM_{\rm cl},
        \label{eqn:totvdp}
\end{equation}
and
\begin{equation}
D_{\rm G}(v_z; \overline{lM_{\rm cl}},\sigma_{\rm lMcl}) = 
         {1\over m_{\rm av}}\,
         \int\limits_{lM_{\rm cl,min}}^{lM_{\rm cl,max}} 
         10^{lM_{\rm cl}}\,\,\xi_{\rm G}(lM_{\rm cl})\,\,
         {\cal V}(v_z,M_{\rm cl})
         \,\, dlM_{\rm cl},
        \label{eqn:totvdg}
\end{equation}
with
\begin{equation}
N = \int\limits_{-\infty}^{+\infty}
    D_{\rm P or G}(v_z)\,dv_z.
\label{eqn:area}
\end{equation}

Equations~\ref{eqn:totvdp} and~\ref{eqn:totvdg} are solved by
numerically integrating $dD_{\rm P}(v_z)/dM_{\rm cl}$ and $dD_{\rm
G}(v_z)/dlM_{\rm cl}$ (initial condition $D_{\rm PorG}(v_z)=0$ for
$M_{\rm cl}=M_{\rm cl,min}=10\,M_\odot$ or $lM_{\rm cl}=lM_{\rm
cl,min}=1$) using a fifth-order Runge-Kutta method with step-size
adaptation (Press et al. 1992) for $1\times10^3$ velocity bins ranging
from $v_{z1}=-500$~pc/Myr to $v_{z2}=+500$~pc/Myr, 500~pc/Myr being
taken as the escape velocity from the solar neighbourhood
(e.g. BT). The mass interval used for the log-normal ICMF, $lM_{\rm
cl,min} = 1$, $lM_{\rm cl,max} = 7$ covers the mass-range of known
star clusters, but implies that the distribution $D_{\rm G}$ needs to
be scaled numerically to $N$, since analytical integration cannot
determine the correct constant $\xi_{\rm 0G}$. Note that the finite
escape velocity from the galaxy implies $\int_{v_{z1}}^{v_{z2}} D_{\rm
PorG}(v_z)\,dv_z < N$ by a negligible amount for ensembles of clusters
ranging up to $M_{\rm cl}\approx10^7\,M_\odot$. Note also that
$lM_{\rm cl,max}=7$ is consistent with the allowed central number
density for the Orion Nebula Cluster (Kroupa 2000).

\begin{figure}
\begin{center}
\rotatebox{0}{\resizebox{0.7 \textwidth}{!}
{\includegraphics{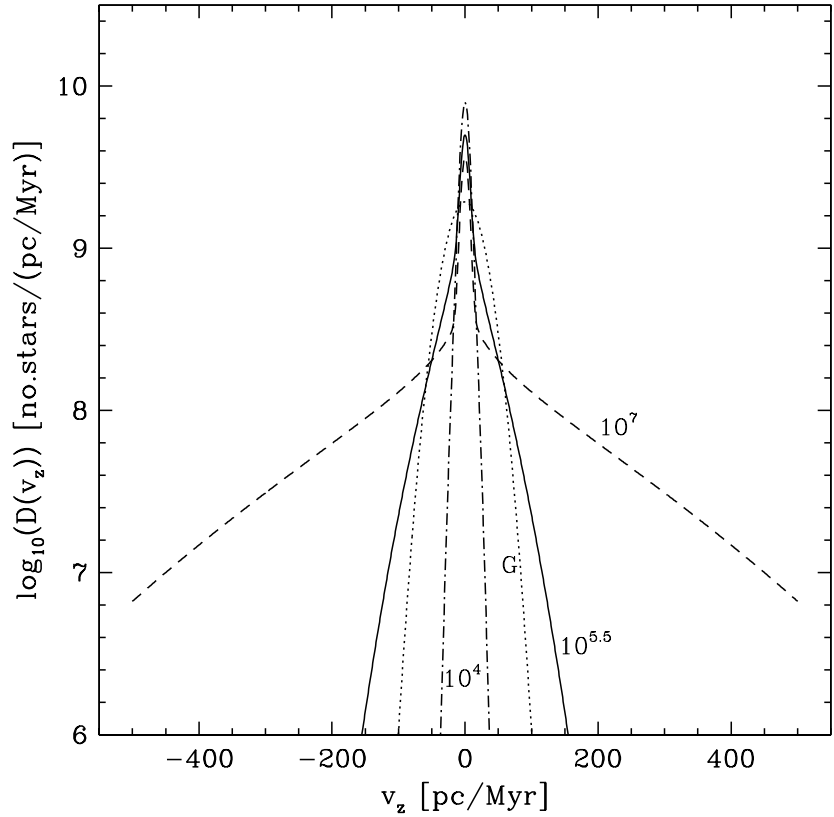}}}
\vskip -27mm
\caption
{\small{Three different velocity distributions, $D_{\rm P}(v_z)$,
assuming the ICMF is a power-law with $\alpha=1.5$ and $M_{\rm
cl,min}=10\,M_\odot$. $M_{\rm cl,max}$ is indicated in $\,M_\odot$
alongside the curves. For the three models (eqns~\ref{eqn:sigmapop_n}
and~\ref{eqn:frac}): $^0\sigma_z=127$~pc/Myr with $\phi=0.38$ ($M_{\rm
cl,max}=10^7\,M_\odot$), $^0\sigma_z=25.8$~pc/Myr with $\phi=0.32$
($M_{\rm cl,max}=10^{5.5}\,M_\odot$), and $^0\sigma_z=6.9$~pc/Myr with
$\phi=0.04$ ($M_{\rm cl,max}=10^4\,M_\odot$). The models assume the
entire Galactic disk of mass $M_{\rm disk}=5\times10^{10}\,M_\odot$ is
buildup of clusters, i.e. $\eta'=1$, so that each distribution has an
area $N=1.25\times10^{11}$~stars (eqn~\ref{eqn:area}).  The thin
dotted curve is a Gaussian, ${\cal G}(\sigma_z=25.8\,{\rm
pc/Myr},\overline{v_z}=0)$, with area $N$.  }}
\label{fig:vdp}
\end{center}
\end{figure}

\begin{figure}
\begin{center}
\rotatebox{0}{\resizebox{0.7 \textwidth}{!}
{\includegraphics{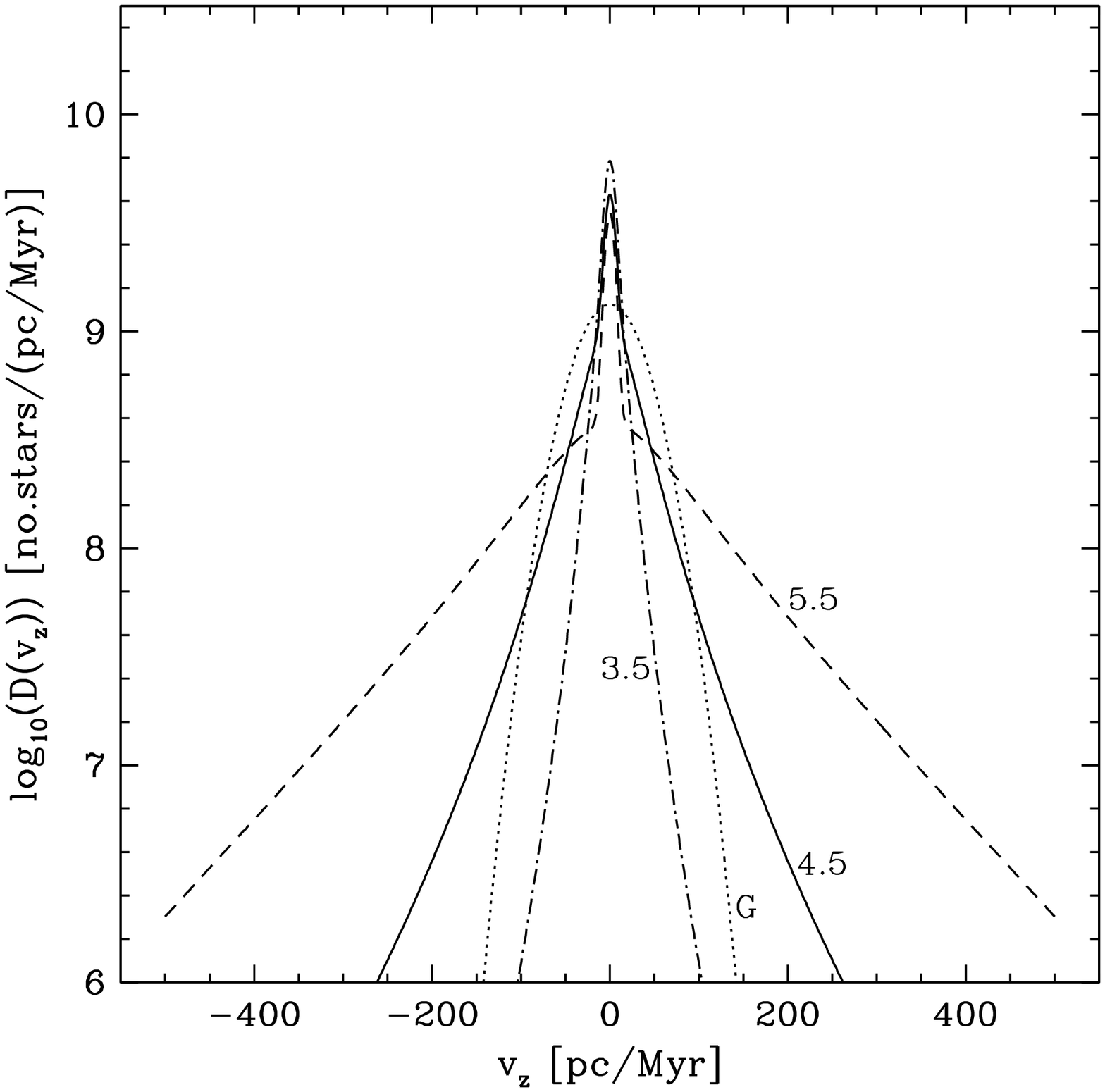}}}
\vskip -27mm
\caption
{\small{As Fig.~\ref{fig:vdp} but assuming the ICMF is a Gaussian in
log$_{10}M_{\rm cl}\equiv lM_{\rm cl}$ with dispersion $\sigma_{\rm
lMcl}=0.5$ and mean $\overline{lM_{\rm cl}}$ shown next to the curves
(mass in $M_\odot$).  For the four models (eqns~\ref{eqn:sigmapop_n}
and~\ref{eqn:frac}): $^0\sigma_z=99.4$~pc/Myr with $\phi=0.33$
($\overline{lM_{\rm cl}}=5.5$), $^0\sigma_z=37.5$~pc/Myr with
$\phi=0.33$ ($\overline{lM_{\rm cl}}=4.5$), and
$^0\sigma_z=12.9$~pc/Myr with $\phi=0.21$ ($\overline{lM_{\rm
cl}}=3.5$).}}
\label{fig:vdg}
\end{center}
\end{figure}

Results for the power-law ICMF with $\alpha=1.5$ are plotted in
Fig.~\ref{fig:vdp}. For Gaussian ICMFs the results are plotted in
Fig.~\ref{fig:vdg}. In general, the resulting velocity distribution,
$D_{\rm PorG}$, is distinctly non-Gaussian.  Note the wings of
high-velocity stars that become more dominant with increasing $M_{\rm
cl,max}$ and $\overline{lM_{\rm cl}}$. The peak at $v_z=0$ comes from
the contribution of $M_{\rm cl}\simless 2000\,M_\odot$ clusters, and
because $\kappa<1$ in eqn~\ref{eqn:truevd}. A cluster--cluster
velocity dispersion, $\sigma_{0z}>5$~pc/Myr, broadens this peak, but
here interest is with the extreme case, namely the kinematical
implications of cluster-formation in a thin gaseous disk.

\subsection{The velocity dispersion and population fraction}
\label{sec:vd}
\begin{figure}
\begin{center}
\rotatebox{0}{\resizebox{0.7 \textwidth}{!}
{\includegraphics{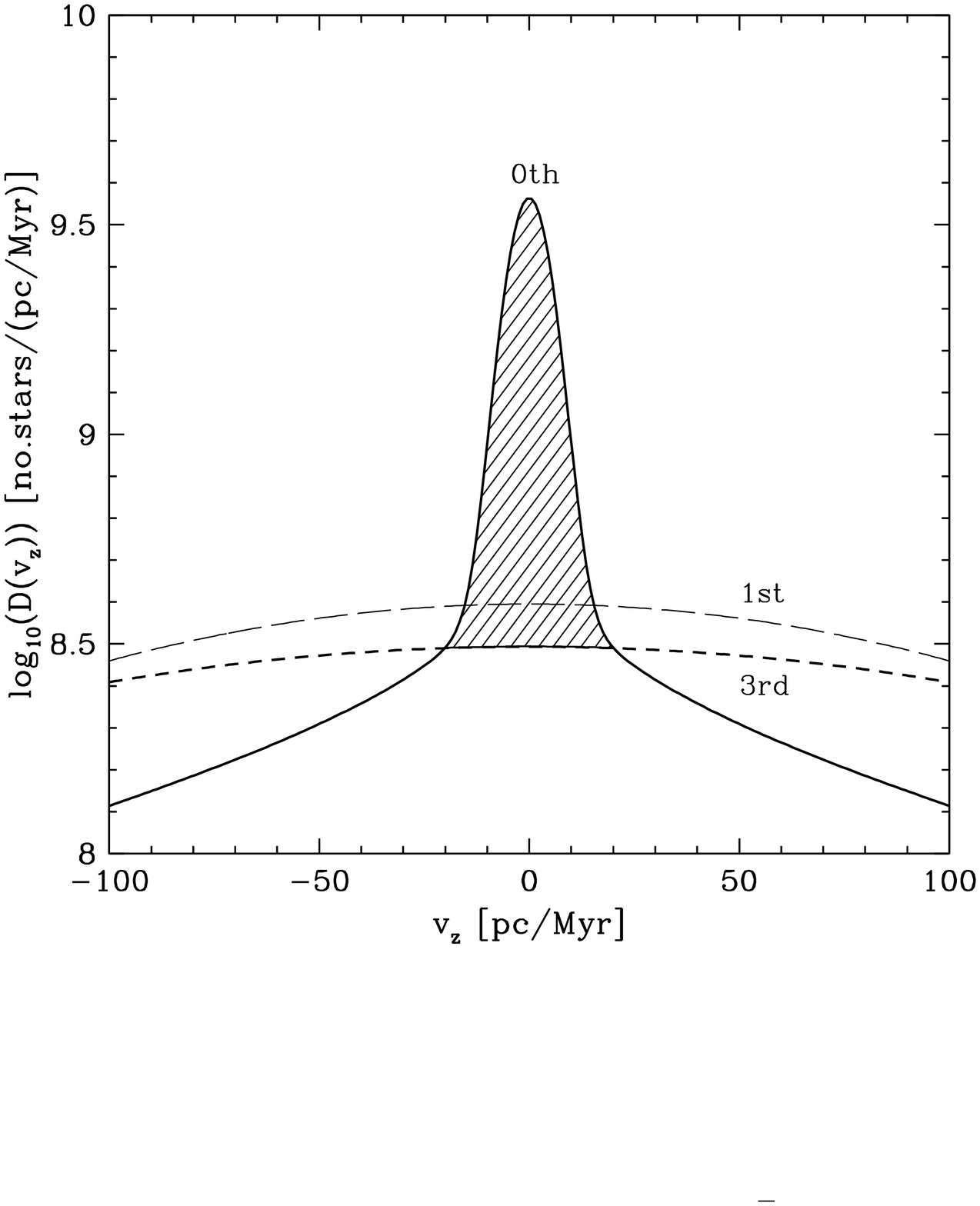}}}
\vskip -27mm
\caption
{\small{Iteration of the velocity distribution function towards an
improved estimate of the velocity dispersion characterising the wings
in the distribution function. In this example the solid curve is
$^0D_{\rm PorG}=D_{\rm P}$ with $M_{\rm cl,max}=10^7\,M_\odot$,
$\alpha=1.5$ and $^0\sigma_z=127$~pc/Myr. The thin long-dashed line is
a Gaussian, ${\cal G}$, with $\sigma_z=127$~pc/Myr. Replacing
$^0D_{\rm P}$ by ${\cal G}$ in the central region where ${\cal
G}<D_{\rm P}$, a first estimate is obtained, $^1D_{\rm P}$, which is
used to recompute the velocity variance
(eqn~\ref{eqn:sigmapop_n}). Two additional such iterations lead to the
corrected distribution $^3D(v_z)$, shown as the lower solid line,
which has the peak (shaded region) removed, and
$\sigma_z=160.1$~pc/Myr (eqn~\ref{eqn:sigmapop}). The shaded area
contains a fraction $\phi=0.38$ of all stars in $^0D_{\rm P}$.  }}
\label{fig:vdp_iter}
\end{center}
\end{figure}

\begin{figure}
\begin{center}
\rotatebox{0}{\resizebox{0.65 \textwidth}{!}
{\includegraphics{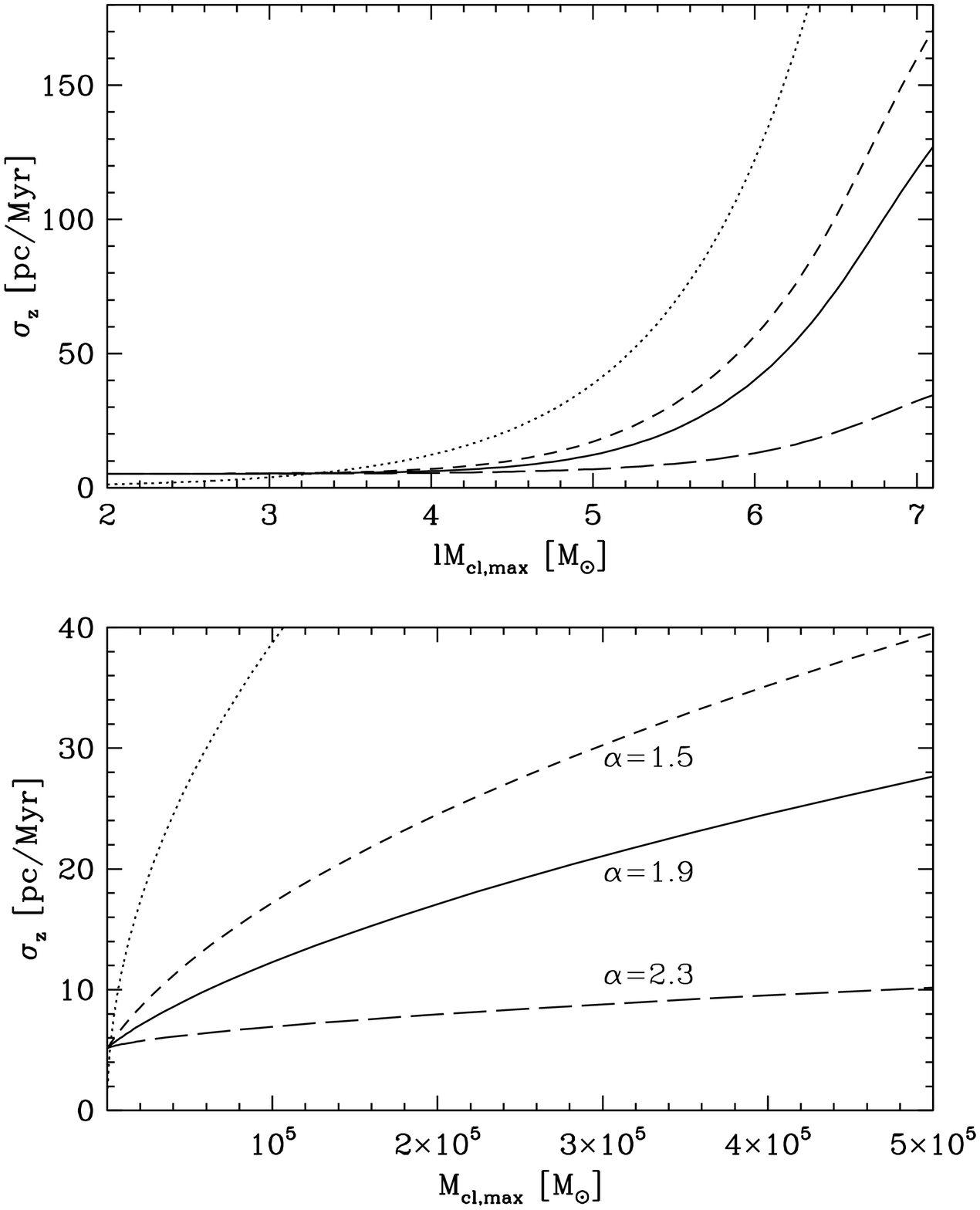}}}
\vskip -5mm
\caption
{\small{Dependence of the velocity dispersion $\sigma_z$
(eqn~\ref{eqn:sigmapop}) on the maximum cluster mass, $M_{\rm
cl,max}$, for a power-law ICMF for three different values of the
power-law index $\alpha$ (line-types indicated in lower panel;
$lM_{\rm cl,max}\equiv {\rm log}_{10}M_{\rm cl,max}$). The dotted line
is the dependence of $\sigma_{0,{\rm cl}}$ on $M_{\rm cl}$ ($=M_{\rm
cl,max}$ in eqn~\ref{eqn:sigma} with $\epsilon R_0=0.3$~pc).  }}
\label{fig:vdisp_p}
\end{center}
\end{figure}

\begin{figure}
\begin{center}
\rotatebox{0}{\resizebox{0.65 \textwidth}{!}
{\includegraphics{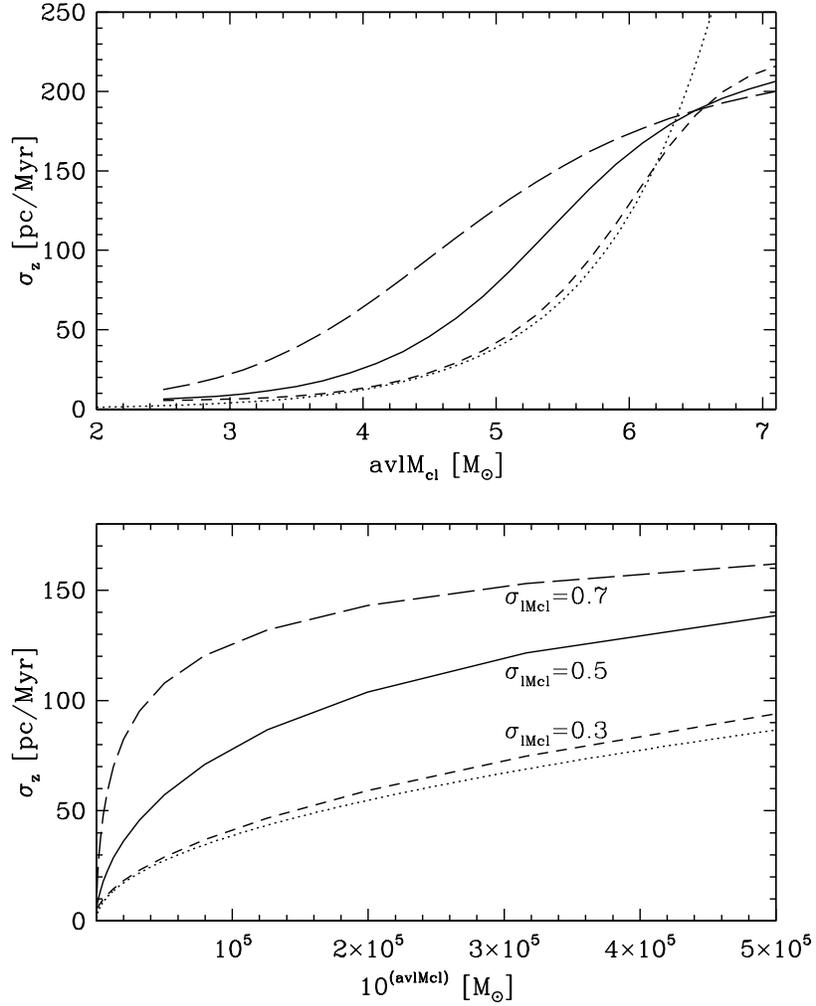}}}
\vskip -5mm
\caption
{\small{Dependence of the velocity dispersion $\sigma_z$
(eqn~\ref{eqn:sigmapop}) on the average log-mass, $avlM_{\rm cl}\equiv
\overline{lM_{\rm cl}}$, for three different values of the standard
deviation in log-mass, $\sigma_{\rm lMcl}$, for a log-normal ICMF
(line-types indicated in lower panel).  The dotted line is the
dependence of $\sigma_{0,{\rm cl}}$ on $M_{\rm cl}$ ($=M_{\rm cl,max}$
in eqn~\ref{eqn:sigma} with $\epsilon R_0=0.3$~pc).  }}
\label{fig:vdisp_g}
\end{center}
\end{figure}

\noindent 
Given $^0D_{\rm P or G}\equiv D_{\rm P or G}$, the variance of the
velocities obtains from
\begin{equation}
^n\sigma_z^2 = {1\over N}
             \int\limits_{v_{z1}}^{v_{z2}} v_z^2 \,
              \left(^nD_{\rm P or G}(v_z)\right)\, dv_z,
\label{eqn:sigmapop_n}
\end{equation}
where $n=0,1,2,3$ are iterations and which, again, is integrated
numerically.  The entire distribution is used to compute the
dispersion $^0\sigma_z$.  An observer would instead try to isolate
distinct sub-populations from the distributions evident in
Fig.~\ref{fig:vdp} or~\ref{fig:vdg} given their non-Gaussian form
(e.g. Reid, Hawley \& Gizis 1995; Gilmore \& Wyse 2001).  This is
modelled here by replacing, in the first iteration, the central,
low-velocity peak in $^0D_{\rm PorG}$ by a Gaussian with dispersion
$^0\sigma_z$, i.e. $^1D_{\rm PorG} = {\cal G}(^0\sigma_z,
\overline{v_z}=0)$ for those $v_z$ near zero for which $^0D_{\rm PorG}
\ge {\cal G}$, but $^1D_{\rm PorG} = \,\, ^0D_{\rm P or G}$
otherwise. This is repeated three times (see Fig.~\ref{fig:vdp_iter}),
until most of the central maximum is removed, yielding the improved
estimate for the velocity dispersion
\begin{equation}
\sigma_z\equiv \,\, ^3\sigma_z.
\label{eqn:sigmapop}
\end{equation}
Figs.~\ref{fig:vdisp_p} and~\ref{fig:vdisp_g} show $\sigma_z$ as a
function of ICMF parameters.

\begin{figure}
\begin{center}
\rotatebox{0}{\resizebox{0.65 \textwidth}{!}
{\includegraphics{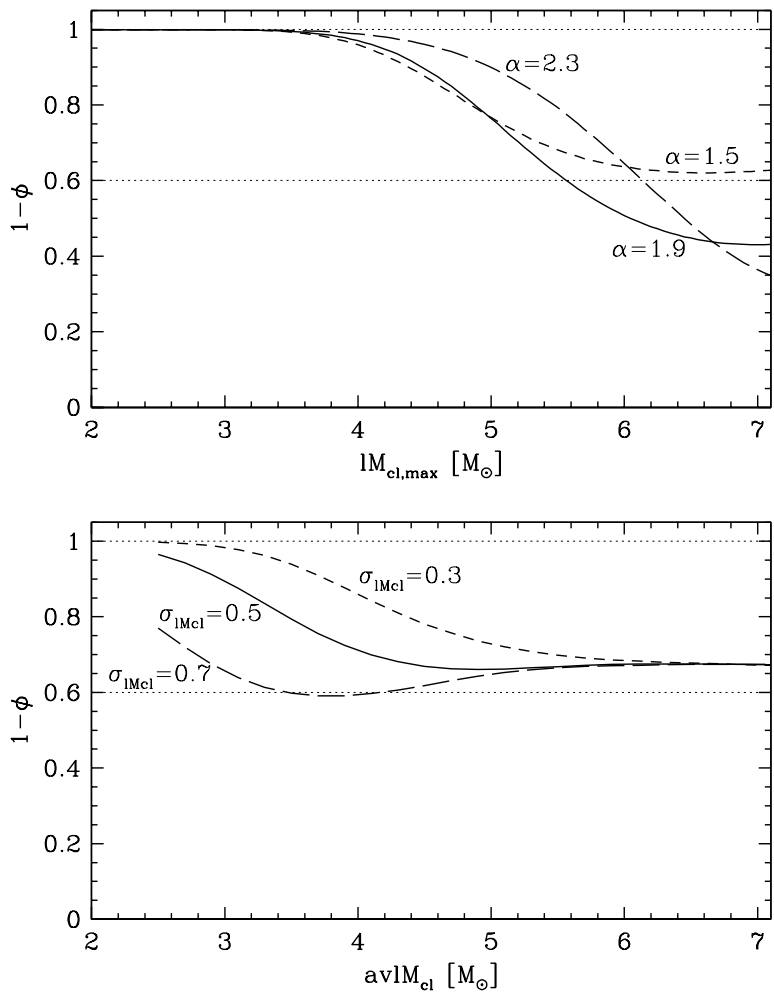}}}
\vskip -5mm
\caption
{\small{ Dependence of the fraction of stars in the wing component
only, $1-\phi$, (eqn~\ref{eqn:frac}) on the maximum cluster mass,
$M_{\rm cl,max}$, for a power-law ICMF for three different values of
the power-law index $\alpha$ (upper panel, as in
Fig.~\ref{fig:vdisp_p}), and on the average log-mass, $avlM_{\rm
cl}\equiv \overline{lM_{\rm cl}}$, for three different values of the
standard deviation in log-mass, $\sigma_{\rm lMcl}$, for a log-normal
ICMF (lower panel, as in Fig.~\ref{fig:vdisp_g}). The region between
the horizontal dotted lines approximately constitutes the range of
acceptable thick-disk models ($1-\phi > 0.6$) provided the constraint
on $\sigma_{\rm z,obs}$ is fulfilled (Fig.~\ref{fig:survad} below). }}
\label{fig:frac}
\end{center}
\end{figure}

An additional constraint is the relative number of stars in the peak
and the whole population,
\begin{equation}
\phi = {^0N - \,\, ^3N \over ^0N},
\label{eqn:frac}
\end{equation}
where $^0N \equiv N$ (eqn~\ref{eqn:area} with $D_{\rm P or G}= \,\,
^0D_{\rm P or G}$), and ${^n N} = \int\limits_{v_{z1}}^{v_{z2}} {^n
D}_{\rm P or G}(v_z)\,dv_z$.  The fraction of the population in the
wings of the velocity distribution is thus $1-\phi$, which is plotted
in Fig.~\ref{fig:frac} for three values of $\alpha$ and $\sigma_{\rm
lMcl}$. Admissible models for the thick disk must not have too many
stars in the peak, so $1-\phi\simgreat 0.6$ is imposed, in addition to
the constraint on $\sigma_z$ (Section~\ref{sec:thickd}).
Note that if the total thick disk amounts to a fraction $\eta'$ of the
mass of the galactic disk, then the observer will identify a fraction
$\eta'\,(1-\phi)$ of the galactic disk as being the actual thick disk,
whereas $\eta'\,\phi$ is the fraction of the thick disk 'hiding'
(subject to adiabatic heating not taken into account here but in
Section~\ref{sec:adheat}) as a thin disk component, but with the same
chemical and age distribution as the observationally identified thick
disk.  Observations suggest
\begin{equation}
\eta=\eta'\,(1-\phi)\approx0.2-0.3
\label{eqn:eta}
\end{equation}
(Section~\ref{sec:intro}) so that $\eta'\simless
0.3-0.5$. In fact, 'thin-disk' stars with [Fe/H]$\,\simless-0.4$
may simply be the low-velocity $\phi$-peak of the thick-disk population.
\subsection{Adiabatic heating}
\label{sec:adheat}
\noindent
If the disk surface mass density, $\Sigma$, of a galaxy accumulates
over time through infall, the velocity dispersion of already existing
stars increases adiabatically leading to an apparent heating of the
disk. Jenkins (1992) finds that adiabatic heating helps to explain the
empirical Wielen-disk-heating in association with scattering on
molecular clouds and spiral arms. The required cloud heating, however,
remains incompatible with the observed molecular cloud properties, so
that the heating agent is still not fully accounted for.

If the time evolution of the infall rate is $\zeta(t) = d\Sigma/dt$
and outflow can be neglected, then
\begin{equation}
\Sigma(t)        = \int_{0}^{t_0} \zeta(t')\,dt' 
                 + \int_{t_0}^{t} \zeta(t')\,dt';
\label{eqn:surfd}
\end{equation}
with $\Sigma_{\rm now} \equiv \Sigma(t_{\rm now})$ and $\Sigma_0
\equiv \Sigma(t_0) \equiv \eta'\Sigma_{\rm now}$
(eqns~\ref{eqn:nstars} and~\ref{eqn:eta}), $t_0\approx4$~Gyr being the
time when the thick disk material with surface mass density $\Sigma_0$
had accumulated (e.g. fig.10.19 in BM; fig.~2 in Fuchs et al. 2001),
and $t_{\rm now}\approx15$~Gyr being the present time. The dependence
on Galactocentric distance, $R_{\rm g}$, is dropped for conciseness,
since the focus here is on the one sample available, namely near the
Sun.  Any infall history $\zeta(t)$ can be envisioned.  Here only a
constant accretion rate onto the thin disk, $\zeta(t)\equiv\zeta =
(\Sigma_{\rm now} - \Sigma_0)/(t_{\rm now}-t_0)$, is considered to
obtain insights into possible implications.

For highly flattened systems and using Jeans equations in the
appropriate limit,
$\Sigma=-1/(2\,\pi\,G\rho)\,\partial{\rho\,\sigma_z^2}/\partial{z}$
(\S4.2 in BT), with the surface density
$\Sigma=\int_{-z}^z\rho(z')\,dz'$, $\rho(z)$ being the mass density at
radial distance $R_g$ from the Galactic centre and height $z$ above
the disk plane.  Assuming the vertical structure of the MW at any
$R_g$ is isothermal ($\sigma_z=\,$const. with $\rho(z)=\rho_0\,{\rm
sech}^2(z/2z_0)\,$) it follows that $\Sigma = (\sigma_z^2/2z_0\pi
G){\rm tanh}(z/2z_0)$.  It can thus be assumed that as the mass of the
disk builds up, $\sigma_z$ of existing stars increases following
$\sigma_z^2 \propto \Sigma$ (see also \S11.3.2 in BM). This implies
\begin{equation}
\sigma_z^2(\Delta t) = \sigma^2_{z,{\rm now}} \, 
               \left( \eta' + \left(1-\eta'\right)\,
              {\Delta t\over \Delta t_{\rm now}} \right),
\label{eqn:veldispad}
\end{equation}
where $\Delta t\equiv t-t_0$ and $\Delta t_{\rm now}\equiv t_{\rm now}
- t_0$.  The velocity spheroid of existing stars, defined by the
velocity dispersions in galactic radial, azimuthal and vertical
directions, distorts and rotates as a consequence of differential
rotation and disk heating. However, the correction terms to $\sigma_z$
are negligible for the present treatment (eqn~23 in Jenkins 1992).

\begin{figure}
\begin{center}
\rotatebox{0}{\resizebox{0.7 \textwidth}{!}
{\includegraphics{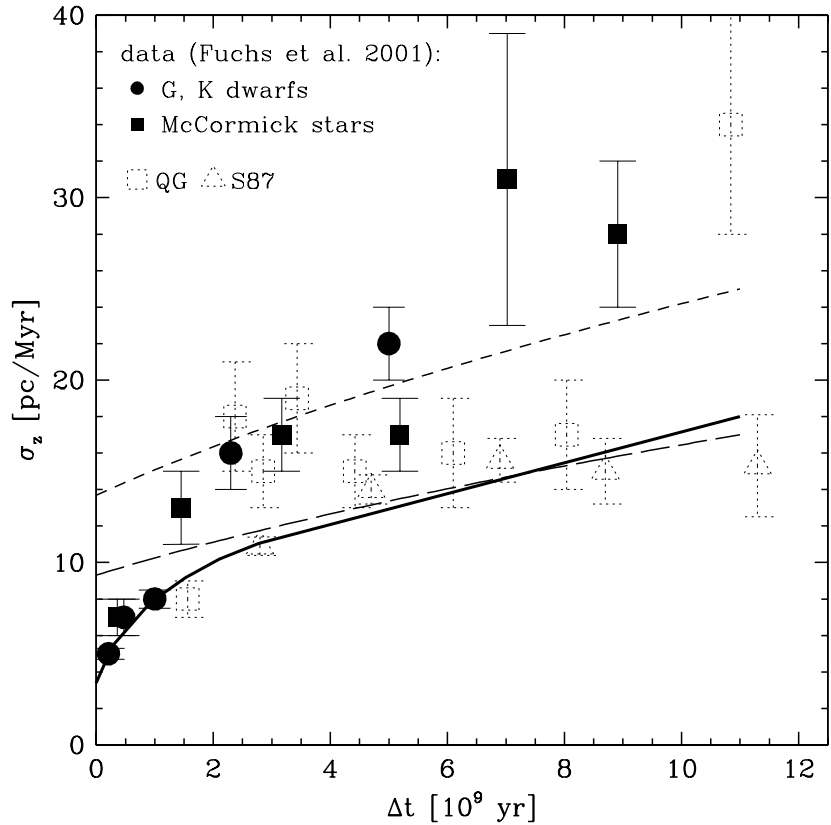}}}
\vskip -27mm
\caption
{\small{The most reliable age--velocity-\-dispersion data are shown as
solid symbols (from fig.~1 in Fuchs et al. 2001).  The short-dashed
and long-dashed lines are eqn~\ref{eqn:veldispad} with $\sigma_{z,{\rm
now}}=25$~and 17~pc/Myr, respectively, assuming that over the past
$\Delta t=11$~Gyr a fraction $1-\eta'=0.7$ of the MW disk assembled
through infall. The solid curve is Jenkins' (1992) model, scaled to
fit the data near $\Delta t=0$ (his fig.~9), young stars having
$\sigma_z\approx 2-5$~pc/Myr (Asiain et al. 1999).  It takes into
account heating through molecular clouds, spiral-wave heating and
adiabatic heating through an accreting disk, and thus incorporates all
known secular heating mechanisms. The dotted symbols indicate
unpublished observational data of Quillen \& Garnett (2000, QG), and
Stroemgren's (1987, S87) data (see Section~\ref{sec:thindat} for more
details). }}
\label{fig:obsvd}
\end{center}
\end{figure}

Assuming the 'thick disk' accumulated within a few~Gyr and formed
30~per cent of the mass now in the MW disk ($\eta'=0.3$),
$\sigma_{0z}\approx22$~pc/Myr at formation ($\Delta t=0$) follows from
the presently observed value $\sigma_{z,{\rm now}}\approx40$~pc/Myr.
Similarly, for the old thin disk $\sigma_{\rm z,obs}\approx25$~pc/Myr
(Fuchs et al. 2001), so that $\sigma_{0z}\approx14$~pc/Myr at star
formation from eqn~\ref{eqn:veldispad}, while the observation of young
stars suggest $\sigma_{0z}\approx2-5$~pc/Myr (Asiain et al. 1999;
Fig.~\ref{fig:obsvd}).

This thus demonstrates that, if the Galactic disk accumulated mass at
a uniform rate over its history, then the required heating of the
thin-disk is reduced, as already shown by Jenkins
(1992). Nevertheless, as evident in Fig.~\ref{fig:obsvd}, additional
heating is required to increase the velocity dispersion within
$\simless3$~Gyr to match the constraints of the Heidelberg group
(Jahreiss et al. 1999; Fuchs et al. 2001).  Jenkins (1992) shows that
molecular clouds and spiral heating cannot account for this 'minimal'
heating (minimal because adiabatic heating is taken into account),
although the possible short-lived nature of molecular clouds are not
taken into account. If these are short lived, with life-times
$\approx10^6-10^7$~yr, then an additional heating source may be active
through the acceleration of young stars as they transgress
time-varying cloud potentials.  Asiain et al. (1999) note that
different young moving groups appear to experience different heating
histories {\it if} in fact their observed velocity dispersion is
assumed to result from the same diffusive mechanism of Wielen
(1977). Alternatively, they suggest that the diffusion constants
differ between associations ('episodic diffusion').

On the other hand, according to the line-of-argument followed here,
individual star-forming events produce associations that expand as a
result of gas loss at different rates, and thus with different
velocity dispersions depending on the particular star clusters sampled
from the ICMF. Expansion thus depends on the intensity of the
star-formation episode.  The history of $\sigma_{z,{\rm obs}}(t)$
plotted in Fig.~\ref{fig:obsvd} may therefore be a mixture of
adiabatic heating, heating from spiral arms and molecular clouds and
fluctuations of the velocity dispersion due to clustered
star-formation.  It will not be easy to disentangle all these effects,
given that the history of clustered star formation is not known, but
an exemplary attempt is made in Section~\ref{sec:thind}. There the
approach is to adopt the age--velocity-dispersion data and Jenkins'
model, and to constrain the ICMFs that are required to give the
unaccounted-for, or 'abnormal', velocity dispersions.

\section{THE POSSIBLE ORIGIN OF THE THICK DISK}
\label{sec:thickd}
\noindent 
Given that clustered star formation is likely to leave a kinematical
imprint along the lines of Section~\ref{sec:veldistr}, the following
question can now be addressed: Since, so far, no satisfying origin for
the thick disk velocity dispersion, $\sigma_{z,{\rm
obs}}\approx40$~pc/Myr, has been found, may it conceivably be
associated with vigorous star-formation during the brief initial
assembly of the thin disk? Observations tell us that violent
molecular-cloud--dynamics is associated with vigorous star formation,
the units of which are star clusters. The maximum cluster mass appears
to correlate with the star-formation rate via the pressure in the
clouds, which in turn is increased when gas clouds are compressed or
collide in perturbed and interacting systems (Elmegreen et al. 2000;
Larsen 2001). So maybe the typical $M_{\rm cl}$ was large when the
thick disk formed?

\begin{figure}
\begin{center}
\rotatebox{0}{\resizebox{0.65 \textwidth}{!}
{\includegraphics{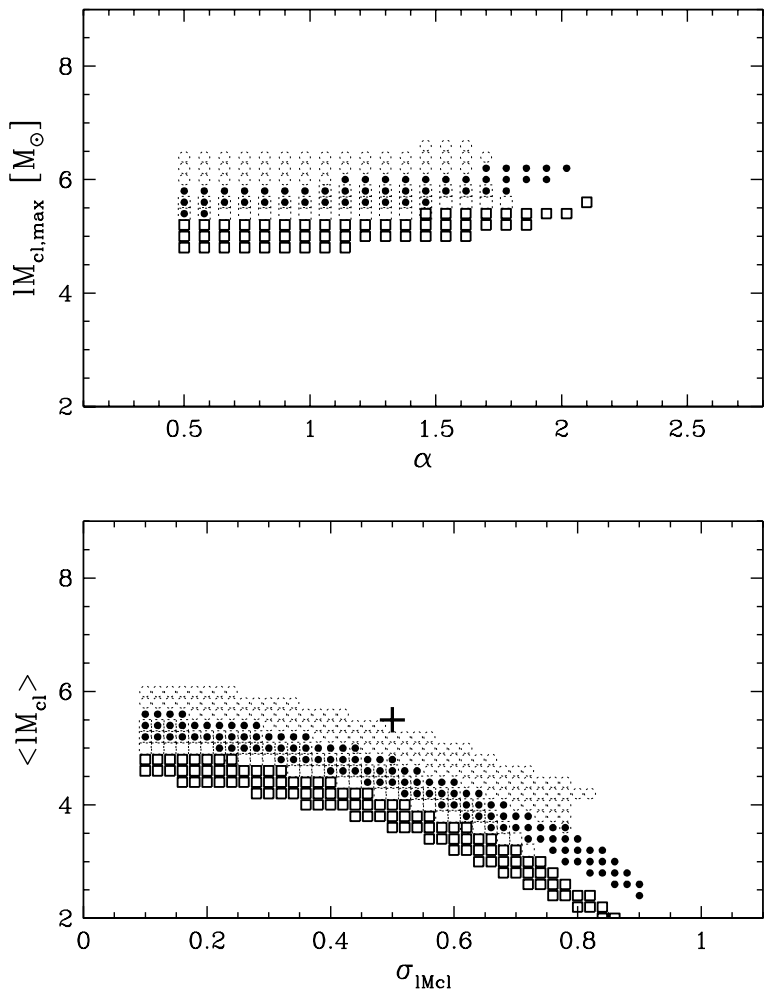}}}
\vskip -5mm
\caption
{\small{{\bf Upper panel:} Solutions in $lM_{\rm cl,max}\equiv{\rm
log}_{10}M_{\rm cl,max},\alpha$ space for star-formation occurring in
clusters distributed as power-law mass functions.  The solution space
indicated as thin dotted squares ($R_0=1$~pc) and thin dotted circles
($R_0=5$~pc) leads to a velocity dispersion, $\sigma_z$, within 30~per
cent of the observed velocity dispersion of the thick Galactic disk,
$\sigma_{z,{\rm obs}}=40$~pc/Myr ($0.7\,\sigma_{z,{\rm obs}}\le
\sigma_z\le 1.3\,\sigma_{z,{\rm obs}}$), whereas thick solid squares
($R_0=1$~pc) and solid circles ($R_0=5$~pc) delineate the solution
space leading to an initial thick-disk velocity dispersion lying in
the 30~per cent interval around 22~pc/Myr ($15.3\le \sigma_z\le
28.5$~pc/Myr), this being the dispersion corrected for adiabatic
heating due to linear growth of the Galactic disk mass after the thick
disk formed. All solutions fulfil $1-\phi>0.6$.  {\bf Lower panel:} As
upper panel, but solutions in $\overline{lM_{\rm cl}}\equiv <lM_{\rm
cl}>, \sigma_{lM_{\rm cl}}$ space for star-formation occurring in
clusters distributed as log-normal mass functions.  The thick cross
locates the ICMF typically inferred for young cluster populations
(Sections~\ref{sec:veldistr}).}}
\label{fig:survad}
\end{center}
\end{figure}

Assuming that the velocity dispersion of the thick disk has not
changed since its formation, the solution for ICMF parameters can be
sought which bring $\sigma_z$ into agreement with $\sigma_{z,{\rm
obs}}$ at the (arbitrary) 30~per cent level. The solutions are
displayed as thin dotted open squares in Fig.~\ref{fig:survad}. The
thick open squares in Fig.~\ref{fig:survad} show the solution space
for the ICMFs leading to $\sigma_z$ within 30~per cent of 22~pc/Myr,
which is the correct value to fit if the thick disk was adiabatically
heated due to a linear growth of the MW disk mass
(Section~\ref{sec:adheat}). The latter solution space is shifted
uniformly to smaller cluster masses by about a factor of five to
ten. If clusters form with $R_0=5$~pc rather than $R_0=1$~pc
(eqns~\ref{eqn:typm} and~\ref{eqn:sigmaexp}), then the corresponding
solution spaces are shifted to larger masses, as is evident in
Fig.~\ref{fig:survad} as the solid and dotted circles for fitting to
the 30~per~cent interval around 22~pc/My and 40~pc/Myr, respectively.

\section{THE THIN DISK HISTORY}
\label{sec:thind}
\noindent 
Returning to the discussion at the end of Section~\ref{sec:adheat},
the solid curve in Fig.~\ref{fig:obsvd} shows the most detailed
available model, $\sigma_{z, {\rm model}}(t)$, by Jenkins (1992) for
disk heating through spiral arms, molecular clouds and the growth of
mass of the MW disk. As stated in the introduction, the model cannot
account for $\sigma_{z, {\rm obs}}(t)$, but it does not take into
account possible short life-times of the molecular clouds
(Section~\ref{sec:adheat}).

\begin{figure}
\begin{center}
\rotatebox{0}{\resizebox{0.7 \textwidth}{!}
{\includegraphics{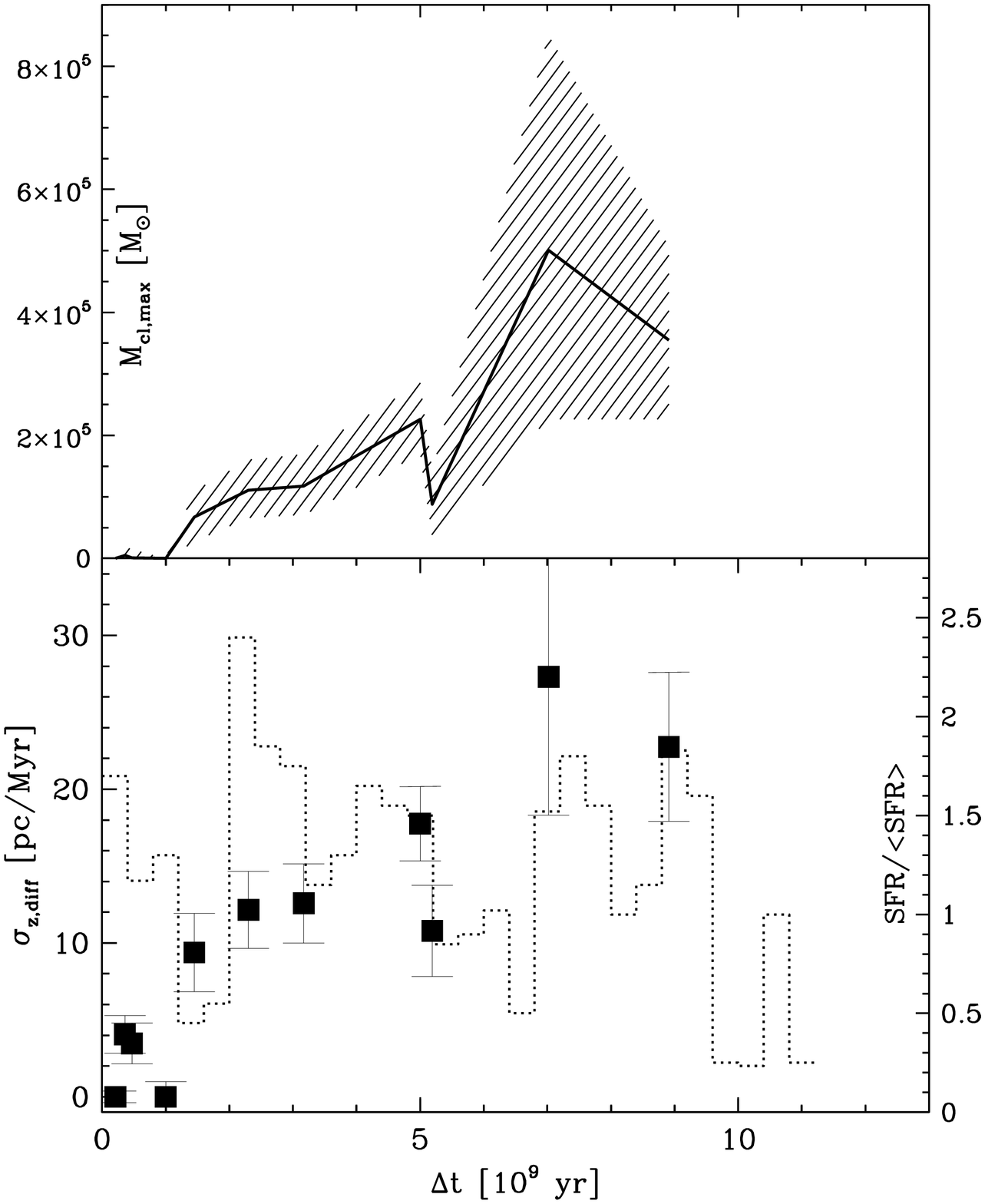}}}
\vskip -0 mm
\caption
{\small{{\bf Bottom panel:} The deviations (eqn.~\ref{eqn:dev}) of
Fuchs et al. (2001) data from Jenkins' (1992) model of the
age--velocity-\-dispersion relation (solid line in
Fig.~\ref{fig:obsvd}) are plotted as solid squares (assuming model has
no errors, the error-bars are $\delta\sigma_{z,{\rm diff}} = \left(
\sigma_{z,{\rm Fuchs}}/\sigma_{z,{\rm diff}} \right)\,
\delta\sigma_{z,{\rm Fuchs}}$).  The star-formation history of the
Galactic disk normalised to the average, SFR/$<$SFR$>$, and as
estimated by Rocha-Pinto et al. (2000), is shown as a dotted
histogram.  {\bf Upper panel:} $M_{\rm cl,max}$ leading to 
$\sigma'_{z,{\rm diff}}$ (eqn~\ref{eqn:devdash}) shown in the lower
panel, assuming a power-law ICMF with $\alpha=1.9$.}}
\label{fig:diffvd}
\end{center}
\end{figure}

A bold step is now taken as an example of how the star-formation
history of the MW disk may be constrained using the approach discussed
here. The residual deviations from the model,
\begin{equation}
\sigma_{z, {\rm diff}}^2(t) = \sigma_{z, {\rm obs}}^2(t) - \sigma_{z,
                      {\rm model}}^2(t)
\label{eqn:dev}
\end{equation}
are plotted in the lower panel of Fig.~\ref{fig:diffvd}.
The deviations, eqn~\ref{eqn:dev}, show an interesting feature in the
age interval 1--4~Gyr and possibly in the age interval 5--7~Gyr. The
former possibly correlates with the burst of star formation, deduced
by Rocha-Pinto et al. (2000) to be at the 99~per cent level over a
constant star-formation rate using chromospheric age estimation, if
the data sets have a relative systematic age difference of about
1~Gyr. This is not entirely unrealistic, given that the age estimators
are indirect (Fuchs et al. 2001). Hernandez, Valls-Gabaud \& Gilmore
(2000) also find significant evidence for an increasing star-formation
rate with increasing age ($\simless 3$~Gyr) using an advanced
Bayesian analysis technique to study the distribution of Hipparcos
stars in the colour--magnitude diagram.  The possible rise in
$\sigma_{z, {\rm obs}}(t)$ near 7~Gyr also finds an increased star
formation rate at $t\simgreat 7$~Gyr when compared to slightly younger
ages, but the significance of this feature is not very high.

Assuming,  for the  purpose  of illustration,  a  power-law ICMF  with
$\alpha=1.9$,  the  upper  panel  of Fig.~\ref{fig:diffvd}  plots  the
maximum cluster mass, $M_{\rm cl,max}$, involved in the star-formation
activity and giving  rise to 
\begin{equation}
\sigma'_{z, {\rm diff}} = \sqrt{\sigma_{z, {\rm  diff}}^2 
                           + \sigma_{0z}^2},
\label{eqn:devdash}
\end{equation}
with $\sigma_{0z}=4$~pc/Myr here; $\sigma'_{z, {\rm diff}}$ is the
correct quantity to fit with $M_{\rm cl,max}$ since eqn~\ref{eqn:dev}
only accounts for the increase of the velocity dispersion over the
current ($\Delta t\approx0$) underlying quiescent star-formation
activity typified by modest embedded clusters, leading to
$\sigma_{0z}\approx4$~pc/Myr.

The analysis above serves to illustrate that clustered star-formation
probably leaves an imprint in the stellar distribution function.
Taking Jenkins' (1992) model to be a correct description of the
secular increase of $\sigma_z$, the above analysis suggests that the
star-formation rate in the MW may have been quietening down, that is,
that the star-formation rate has been declining, and/or that the ICMF
may have been changing towards smaller cluster masses. 

\section{DISCUSSION}
\label{sec:disc}
\noindent 
The analysis presented in this contribution serves to demonstrate that
the kinematics of stars in the MW disk and other galaxies may be
influenced by stellar birth in clusters.  However, the application to
specific kinematical abnormalities in the thin disk, and to the thick
disk, is to be viewed as suggestive rather than definitive.

\subsection{The model}
The model set-up in Section~\ref{sec:veldistr} is simple but
sufficient for this first exploration of how the kinematically hot
component emerging from a cluster-forming event may affect the stellar
distribution function in a galaxy.

The model (eqn~\ref{eqn:truevd}) neglects possible variations of the
star-formation efficiency with cluster mass and the possible variation
of the gas-expulsion time-scale, which may be much longer than the
dynamical time of embedded massive clusters (Tan \& McKee 2001). If
this is so, then the stellar cluster reacts adiabatically to gas
removal, and the final velocity dispersion in the cluster will be
reduced by a factor $\epsilon$ (Mathieu 1983). The velocity
distribution function of escaping stars is, in this case, much more
difficult to estimate. It nevertheless remains a function of $M_{\rm
cl}$, so that the essence of the argument here remains valid, although
the detailed solutions evident in Figs.~\ref{fig:survad}
and~\ref{fig:diffvd} would change somewhat towards larger cluster
masses.  Furthermore, the present models do not take into account that
the star clusters, which form as nuclei of the expanding stellar
associations, continue to add to the thick disk by ejecting
high-velocity stars due to three- and four-body encounters. This is
evident for massive young clusters (figs.~2 and~3 in Kroupa 2001c) as
well as clusters that typically form the Galactic field population
(Kroupa 1998).  While such events are rare, the implication is that
the hot disk component should have a complex metallicity and age
structure. A next stage in this project is to assemble synthetic disk
populations using high-precision direct $N$-body calculations of
cluster ensembles.

This model predicts a complex velocity field in galaxies that are
actively forming star-clusters. In the MW this is evident through
expanding OB associations, and the empirical finding that the velocity
dispersions of theses differ (Asiain et al. 1999). OB associations are
most probably only the central regions of the expansive flow, because
massive stars either form centrally concentrated, or they sink to the
centres of their embedded cluster essentially on a dynamical
timescale. If a young cluster can approximately establish energy
equipartition between the stars before gas expulsion, then the massive
sub-population will have a significantly smaller velocity dispersion
than the low-mass members. Furthermore, an expanding association
'looses' its fast-moving stars early. An observer with a limited
survey volume therefore underestimates the velocity dispersion.

The exploratory analysis presented here neglects the other velocity
components. Focus is entirely on the component perpendicular to the
Galactic disk, easing the analysis. However, according to the notion
presented here, the birth of a star cluster leads to an initially
isotropically expanding stellar population. The velocity spheroid,
given by the ratio of the velocity dispersion perpendicular to the
disk ($\sigma_W=\sigma_z$), towards the Galactic centre ($\sigma_U$)
and in the direction of the motion of the local standard of rest
($\sigma_V$), is thus initially spherical. Stars expelled from their
cluster towards the Galactic centre accelerate and overtake the other
stars, while stars leaving the cluster in the opposite direction are
decelerated by the Galactic potential and fall behind. The velocity
spheroid is thus distorted due to the differential rotation, and it
will be an interesting problem to investigate if the observed velocity
spheroid is consistent with the present notion.  For the thin disk
Fuchs et al. (2001) gives $\sigma_U \,:\, \sigma_V \,:\, \sigma_W =
3.5 \,:\,1.5 \,:\, 1$, independent of age, while Chiba \& Beers (2000)
find for the thick disk $\sigma_U \,:\, \sigma_V \,:\, \sigma_W = 1.31
\,:\,1.43 \,:\, 1$. Such differences may partially be due to secular
heating mechanisms such as scattering off molecular clouds and spiral
arms that decrease $\sigma_W/\sigma_U$ (Gerssen, Kuijken \& Merrifield
2000). This is more efficient for the colder thin disk
population. Adiabatic heating due to the growth of the MW disk also
evolves the velocity spheroid, and a passing satellite galaxy induces
a radial compression of the gaseous disk which can lead to enhanced
star formation and an increased radial velocity dispersion owing to
star formation in the radially inflowing gas and the formation of a
bar.  A full model of the velocity spheroid is thus highly non-trivial
and subject to the details of the satellite orbit and the evolution of
the Galaxy.

\subsection{The thin disk}
The literature contains various estimates of
age--velocity-dispersion data, and contradictory claims have been
made. It is thus useful to cast an independent eye at the situation
(Section~\ref{sec:thindat}). The star-formation history of the
thin-disk is also briefly discussed in Section~\ref{sec:thinhist}. 

\subsubsection{Data}
\label{sec:thindat}
The most reliable age--velocity-dispersion data come from the
(distance-limited) solar-neighbourhood sample contained in the fourth
version of the catalogue of nearby stars originally compiled at the
Astronomisches Rechen-Institut (ARI) in Heidelberg by the late Gliese
and continued by Jahreiss (CNS4, e.g. Jahreiss et al. 1999; Fuchs et
al. 2001). This sample has been yielding consistent results over the
past 2--3 decades (e.g. Wielen 1977) despite significant improvements
such as the inclusion of Hipparcos data for the majority of the
stars. The data are shown in Fig.~\ref{fig:obsvd}.

Since theoretical work has not been able to explain the strong
apparent heating of stars with age, the data set has been questioned
by a variety of researchers. For example, BM point to the work of
Stroemgren (1987). Stroemgren used a preliminary version of the
Edvardsson, Andersen \& Gustafsson (1993) catalogue to select a sample
of stars with $-0.15\le$[Fe/H]$\le+0.15$ and found essentially no
increase in $\sigma$ for ages older than about 4~Gyr, in stark
contrast to the results based on the CNS4 data, suggesting that this
metal-rich thin-disk subpopulation was not subject to the heating
mechanism advocated by Wielen and co-workers at the ARI. From
Fig.~\ref{fig:obsvd} it in fact follows that the best available
theoretical work is in next to excellent agreement with these data.
Such a situation may arise if the local stellar sample is contaminated
by thick-disk stars, but this would require some thick disk stars to
be mis-classified as being significantly younger than the thick-disk
bulk. Alternatively, the metal-rich thin-disk population may have been
born in modest embedded clusters leading to no kinematical
abnormality. Modern data should be used to verify if metal-rich and
metal-poor thin-disk populations do indeed follow different
age--velocity-dispersion relations.

An additional age--velocity-dispersion data set that is not consistent
with the CNS4 data has been submitted for publication by Quillen \&
Garnett (2000). Again, this data set, which is a re-analysis of the
Edvardsson et al. (1993) data, is in perfect agreement with Jenkins'
model and with the data of Stroemgren (1987), but also shows the
kinematical peculiarity in the age-interval 1--4~Gyr evident in the
CNS4 data (Fig.~\ref{fig:obsvd}).

While the Stroemgren (1987) data are interesting, no follow-up work
appeared in a refereed journal. Also, the Quillen \& Garnett (2000)
paper will not be published (Quillen, private communication), and
Fuchs et al. (2001) demonstrate that their own analysis of the
Edvardsson et al. (1993) data yields essentially the same results as
they obtain from the CNS4 (fig.~2 in Fuchs et al. 2001). Furthermore,
the large velocity dispersions of the older CNS4 thin-disk stars are
verified by independent work. Notably, Reid et al. (1995, their
table~6) also find $\sigma_z\approx 25$~pc/Myr for stars with
$8<M_V<15$. Such a large $\sigma_z$ is also obtained for stars within
the very immediate solar-neighbourhood with distances less than 5.2~pc
(Kroupa, Tout \& Gilmore 1993). Finally, Binney, Dehnen \& Bertelli
(2000) find, from a statistical analysis of their own new sample of
kinematical data essentially the same age--velocity-dispersion
relation as advocated by Fuchs et al. (2001, their fig.~4).

The above discussion serves to demonstrate on the one hand that some
uncertainties remain in the definition of the age--velocity-dispersion
data, but that the evidence lies in favour of the results obtained
from the CNS4 (e.g. Fuchs et al. 2001), i.e. that stars have velocity
dispersions that increase more steeply with age than theory can
account for.

\subsubsection{On its history}
\label{sec:thinhist}
It is notable that the suggestive finding in Section~\ref{sec:thind}
that the star-formation rate may have declined in the last few~Gyr is
similar to the conclusion of Fuchs et al. (2001) based on the
cumulative star-formation rate as traced by a sample of~G and~K
solar-neighbourhood dwarfs with measured chromospheric emission fluxes
(their fig.~5). According to these data, the MW thin disk transformed
into a quiescent star-formation mode about 3~Gyr ago. Hernandez et
al. (2000) also find a decreasing star-formation rate during the
past~3~Gyr (Section~\ref{sec:thind}).  Given the good empirical
correlation between the mass of the most massive cluster formed and
the star-formation rate documented by Larsen (2001) for a sample of
22~galaxies, it would thus not be surprising to find that the
star-formation history of the MW disk is reflected in variations of
the ICMF, as suggested in the upper panel of Fig.~\ref{fig:diffvd}. 

Janes \& Phelps (1994) indeed find an old cluster population of
Galactic clusters that have a vertical scale-height of about 300~pc
similar to old thin disk stars, and may thus have been heated from
$\sigma_{z,0}\approx5$~pc/Myr to about 15~pc/Myr as described by
Jenkins' model (Fig.~\ref{fig:obsvd}), although cluster-survival of
scattering events is controversial, and it may be necessary to allow
$\sigma_{0z}>5$~pc/Myr during the star-burst that formed this cluster
population.  These clusters are confined to Galactocentric radii
$\simgreat7.5$~kpc being qualitatively consistent with the present
MW-disk perturbation hypothesis. Janes \& Phelps identify this cluster
population with a star-burst about 5--7~Gyr ago, which corresponds to
the maximum in the ICMF at about that time as inferred here
(Fig.~\ref{fig:diffvd}).

\subsection{The thick disk}
The thick-disk ICMF constrained in Section~\ref{sec:thickd} implies
that star-clusters may have formed with masses extending to
$10^{5-6}\,M_\odot$. Some of these may have survived to the present
day.  If it is correct that thick-disk star-formation occurred in a
thin gaseous disk, as assumed in Section~\ref{sec:thickd}
($\sigma_{0z}=5$~pc/Myr), then the surviving population of clusters
ought to now have a vertical velocity dispersion of about 20~pc/Myr,
according to Jenkins' model (Fig.~\ref{fig:obsvd}).

The extreme assumption $\sigma_{0z}=5$~pc/Myr was made merely
to study if the hot kinematical stellar component emerging from
star-cluster birth can give rise to the thick-disk kinematics for
plausible ICMFs. Here the view is that a quiescent thin gas-rich MW disk may
have assembled early-on, the MW then essentially being a low-surface
brightness disk galaxy. A perturbation by a passing satellite galaxy
may have induced significant star formation leading to the thick disk
component. 

However, it may well be that the thick disk formed with a
cluster--cluster velocity dispersion $\sigma_{0z}\approx 22$~pc/Myr
due to the probably turbulent and chaotic conditions leading to the
assembly of the thin disk, later being heated adiabatically
to~35~pc/Myr. In this alternative and maybe more plausible scenario,
the resulting ICMF solutions would be only slightly shifted to lower
cluster masses (Fig.~\ref{fig:survad}).  Fossils of this vigorous
star-formation event with $\sigma_{0z}\approx22$~pc/Myr may be the
metal-rich MW globular cluster system which has a density distribution
very similar to the stellar thick disk (e.g. BM).

According to Minniti (1996) and Forbes, Brodie \& Larsen (2001) a
large fraction of these metal-rich globular clusters may rather belong
to a bulge population, so that a definitive identification of
metal-rich globular clusters as being the fossils of the thick-disk
star-formation events cannot be made. Any such population of clusters
is likely to have suffered some orbital decay towards the MW disk
through non-isotropic dynamical friction on the disk (Binney 1977),
implying that the original thickness of the metal-rich globular
cluster disk population may have been larger than the present value,
possibly compromising its association with the stellar thick disk.

\section{CONCLUSIONS}
\label{sec:concs}
\noindent 
The age--velocity-\-dispersion relation for the MW is addressed by
considering the effects that star-cluster formation may have on the
velocity field of stars.

Star clusters form with star-formation efficiencies $\epsilon<0.4$ and
they are probably near global gravitational equilibrium just before
the remaining gas is expelled. Observational evidence suggests that
gas expulsion is rapid, so that the majority of stars expand outwards
with a velocity dispersion that is related to the velocity dispersion
prior to gas expulsion. 

This scenario implies that the distribution function of stars in
galaxies should reflect these events. Peculiar kinematical features,
such as enhanced velocity dispersions for stars of particular ages,
may thus be the result of specific star-formation events, or bursts,
that produced star-clusters with masses larger than that typical under
more quiescent star-forming conditions. An empirical correlation
between cluster mass and star-formation rate is established by
Larsen (2001).

The calculations presented here show that the velocity distribution
function of stars emerging from a star-formation event is distinctly
non-Gaussian with a low-dispersion $\phi$-peak and broad wings
(Figs.~\ref{fig:vdp} and~\ref{fig:vdg}). It is noteworthy that Reid et
al. (1995) find that the velocity distribution of solar-neighbourhood
stars shows structure reminiscent of the present models. Ancient
thin-disk stars with overlapping metallicities to those of thick-disk
stars may, according to this interpretation, be the kinematically cold
$\phi$-peak of the star-formation burst that formed the thick disk.

Examples of star-formation bursts leading to abnormal kinematical
signatures may be the Galactic thick disk, and possible star-formation
bursts about 1--4~Gyr and 5--7~Gyr ago evident in the
age--velocity-\-dispersion data for the solar neighbourhood. These
kinematical 'abnormalities' are analysed here by calculating the
velocity distribution function assuming stars form in ensembles of
star clusters distributed either as power-law or as log-normal initial
cluster mass functions (ICMFs). The allowed ICMFs are constrained
(Figs.~\ref{fig:survad} and~\ref{fig:diffvd}). The results are
plausible when compared to direct surveys of the ICMF
(Section~\ref{sec:veldistr}).  The putative star-formation burst
5--7~Gyr ago may be associated with the population of old (and thus
initially relatively massive) Galactic clusters identified by Janes \&
Phelps (1994).

Satellite infall or perturbation may still be responsible for a burst
of star formation that allows more massive clusters to form, but, as
shown here, heating of galactic disks may not require direct satellite
infall into the disk. It is thus interesting that Rocha-Pinto et
al. (2000) find possible correlations between their star-formation
history and the orbit of the Magellanic Clouds, which in turn can be
correlated with the above mentioned age--velocity-\-dispersion
abnormalities (Fig.~\ref{fig:diffvd}). Schwarzkopf \& Dettmar (2000;
2001) find that perturbed disk galaxies are significantly thicker than
isolated disk galaxies. The present notion does not exclude that disk
thickening can also be induced through satellites or high-velocity
clouds hitting disks directly.

The notion raised here readily carries over to populating the Galactic
spheroid through forming star clusters, only the most massive of which
survived to become the present-day globulars. It also carries over to
properties of the distribution function of moving groups associated
with the formation of individual distant globular clusters in the
out-reaches of the Galactic halo.

The concept introduced here relies on the unbound star-cluster
population expanding with a velocity that depends on the cluster
configuration before gas expulsion (eqn.~\ref{eqn:truevd}).  This can
be tested with high-precision astrometry. A perfect test-bed is the
Large Magellanic Cloud which is forming star clusters profusely,
probably as a result of being tidally perturbed (van der Marel
2001). Thus, upcoming space-astrometry missions (DIVA: R\"oser 1999;
GAIA: Lindegren \& Perryman 1996 and Gilmore et al. 1998) should
measure complex velocity fields in the LMC, since each very young but
gas-free cluster should have associated with it an approximately
radially expanding population of stars with an overall velocity
dispersion characteristic of the conditions before
gas-expulsion. Evidence for this is emerging
(Section~\ref{sec:csf}). If an empirical upper limit on the velocity
dispersion of expanding associations independent of $M_{\rm cl}$ for
some value larger than $M_{\rm cl,crit}$ can be found, then this would
serve as an important constraint on the star-formation efficiency and
gas-expulsion time for clusters more massive than $M_{\rm cl,crit}$.
It will also be necessary to quantify the evolution of the velocity
spheroid ($\sigma_U \,:\, \sigma_V \,:\, \sigma_W$) as an improved
test of the model against available observational constraints.

\acknowledgements 
\vskip 10mm
\noindent{\bf Acknowledgements}
\vskip 3mm
\noindent 
I thank Christian Theis for friendly discussions.

%

\end{document}